\documentclass[a4paper,twoside]{article}
\usepackage{hyperref}
\usepackage{epsfig}
\usepackage{subcaption}
\usepackage{calc}
\usepackage{amssymb}
\usepackage{amstext}
\usepackage{amsmath}
\usepackage{amsthm}
\usepackage{multicol}
\usepackage{pslatex}
\usepackage{apalike}
\usepackage{SCITEPRESS}  
\usepackage{graphicx}
\usepackage{float}
\usepackage{xcolor}
\usepackage{ragged2e}
\usepackage{siunitx}
\usepackage{indentfirst}

\begin{document}

\title{Model-based analysis support for dependable complex systems in CHESS}

\author{\authorname{Felicien Ihirwe\sup{1}, Silvia Mazzini\sup{1}, Pierluigi Pierini\sup{1}, Stefano Tonetta\sup{2} and Alberto Debiasi\sup{2}}
\affiliation{\sup{1}Innovation Technology Service Lab, Intecs Solution Spa, Pisa, Italy}
\affiliation{\sup{2}Embedded Systems Unit, Fondazione Bruno Kessler (FBK), Povo, Italy}
\email{\{felicien.ihirwe,silvia.mazzini,pierluigi.pierini\}@intecs.it, \{adebiasi,tonettas\}@fbk.eu}
}



\abstract{
The challenges related to dependable complex systems are heterogeneous and involve different aspects of the system. On one hand, the decision-making processes need to take into account many options. On the other hand, the design of the system logical architecture must consider various dependability concerns such as safety, reliability, and security. Moreover, in case of high-assurance systems, the analysis of such concerns must be performed with rigorous methods. In this paper, we present the new development of CHESS, a cross-domain, model-driven, component-based and open-source tool for the development of high-integrity systems. We focus on the new recently distributed version of CHESS, which supports extended model-based development and analyses for safety and security concerns. Finally, we present contributions of CHESS to several international research projects.}

\maketitle

\section{\uppercase{Introduction}}

The ever increasing complexity and dependability issues of systems in various domains, such as transportation, space, energy, health, and industrial production, requires effective design and development methods. The complexity and heterogeneity of components requires modeling approaches that spans different technical disciplines and prove effective in the end-to-end engineering of the products. This imply taking into account various requirements such as quality, performance, cost, safety, security, and reliability. Model-based design technologies enable the user to perform beforehand different assurance related activities such as physical architecture exploration, system's behavioral analysis,  early verification, and validation.

CHESS \cite{chessAid} toolset offers a cross-domain modeling and analysis of high-integrity systems providing an integrated framework that helps the modeler (user) to automate different development phases: from the requirements definition, to the architectural modeling of the system’s software and hardware, up to its deployment to hardware platform \cite{chessAid}. CHESS follows a component-based approach where the user decouple different functional parts of the system as components that can be modeled, analyzed, verified, stored, reused individually, and be integrated to meet the system's common goal. 
CHESS supports, among other, schedulability and dependability analysis across the entire projects life cycle. The results of the analysis are back-propagated to the model itself so that the modeler can review and fine-tune the model to satisfy real-time and dependability requirements \cite{chesstoolguide}. 

CHESS recently became a full-fledged open-source project, hosted by The Eclipse Foundation (\url{https://www.eclipse.org/chess/}). The code is developed by various contributors following an open-source approach with public projects for issue tracking, code repository branches, and continuous integration. This paper presents the latest development of CHESS to address software's security and safety analysis of the system exploiting the CHESS error model to represent faults and attacks. Analysis is completed by the integration with back-end tools such as xSAP~\cite{BittnerBCCGGMMZ16} and Mobius~\cite{mobius} for the minimal cut sets analysis and Monte-Carlo simulation.

We report on various projects that used CHESS to provide evidence for the assurance of complex systems and on the functionalities that have been added for such purpose. So far CHESS has been applied in domains such as Avionics \cite{chessavionic}, Automotive \cite{chessautomotive}, Space \cite{chessVerification}, Telecommunication \cite{chesstelecom}, and Petroleum \cite{chesspetroleum}\cite{FTAandFMEA}. 

The rest of the paper is arranged into six sections. Sections \ref{sec:chessInNut} provides an introduction to the CHESS tool building blocks and methodology, Section \ref{sec:updates} presents the new major features released under CHESS 1.0.0, Section \ref{sec:chessInPractice} presents some practical applications on different projects, Section \ref{sec:relatedWork} discusses the related work with respect to CHESS approach. In Section \ref{sec:futureWork} we present the envisioned future extension on CHESS and finally Section \ref{sec:conclusion} concludes the paper.

\section{\uppercase{CHESS IN A NUTSHELL}} \label{sec:chessInNut}
\begin{figure}[!h]
  \centering
   {\epsfig{file = 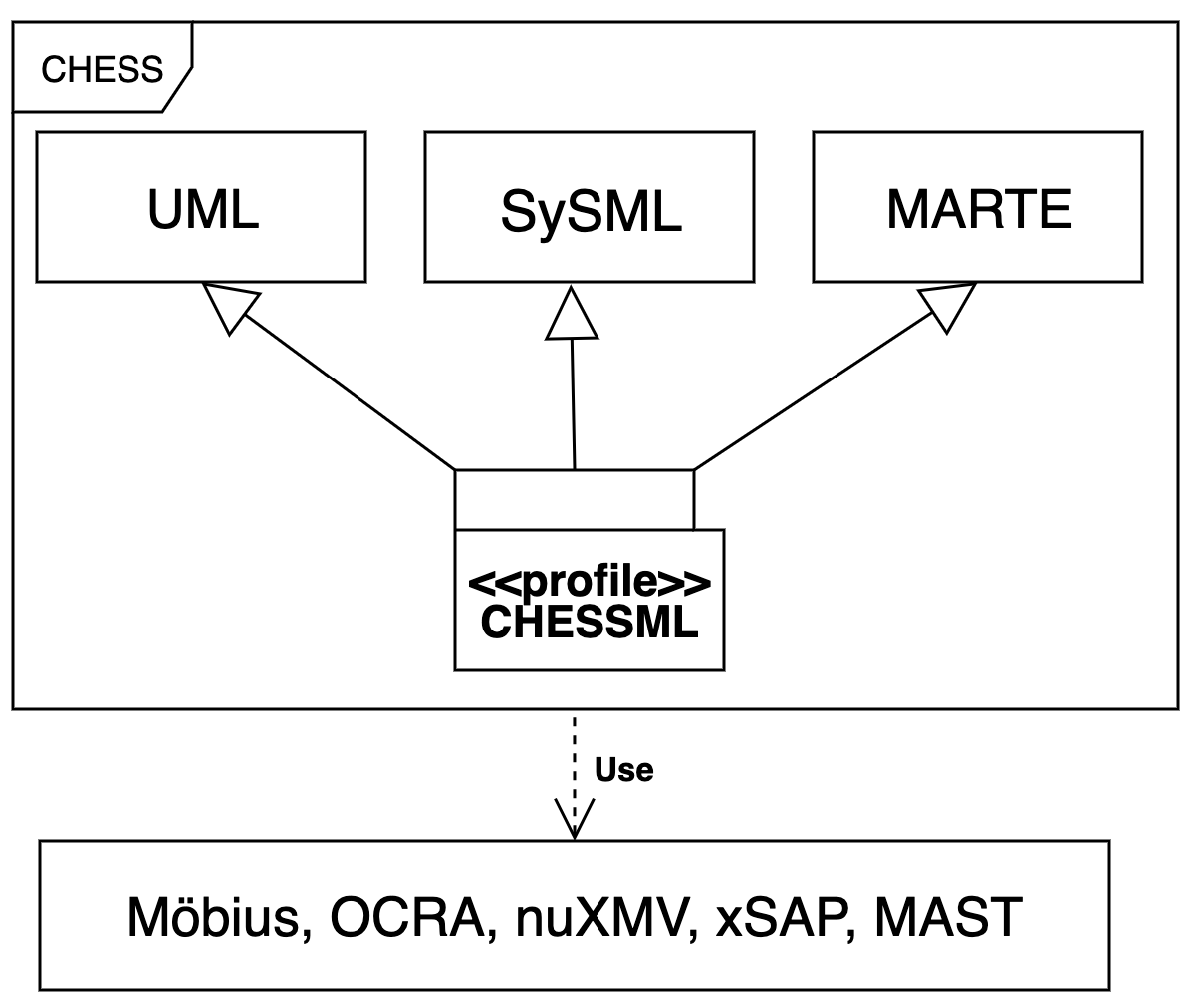, width = 7cm}}
  \caption{CHESS high level structure}
  \label{fig:chessStructure}
 \end{figure}

The CHESS modeling tool was released under the Eclipse PolarSys project\footnote{\url{https://projects.eclipse.org/projects/polarsys.chess}} and recently it was moved from the incubation status to the first major release. CHESSML is an integrated modelling language profiled from OMG standard languages: UML, SysML and MARTE under the Papyrus modeling environment \cite{papyrus}. Figure \ref{fig:chessStructure} shows the high-level architecture of CHESS infrastructure. Not all the features from all the three languages were used, CHESS only exploits specific subsets of the them that suits its perspective. There are different tools, plugins, and languages that were integrated into CHESS to support model validation, model checking, realtime and dependability analysis. 

In this section, we are going to look at different core aspects of CHESS methodology, development features, and implementation mechanisms.  We will also look in depth at different analysis mechanisms that are being performed in CHESS and how they link together to enhance system correctness.

\subsection{Component-based methodology}
CHESSML language supports a component-based development methodology. Emphasis is given to \textit{separation of concerns} between the functional and the non-functional dimensions, such as safety, security, reliability, performance, and robustness \cite{chessIoT}.

In CHESS, components at design level encompass functional concerns only (i.e., they are devoid of any constructs pertaining to tasking and specific computational model). The specification of non-functional attributes is then used for the automated generation of the container, enforcing the realization of the non-functional attributes declared for the component to be wrapped.

This CHESS component-based approach enforces
two main aspects of modeling: “Compositionality”
and “Composability”. Compositionality implies that all the properties of the whole system are determined as the collection of the properties of the constituting components and its execution environment. Composability is achieved when the individual components’ properties are preserved from its definition, development, and deployment on the target platform \cite{chessIoT}. These two component’s properties are conceived through the whole development process to support the “composition with guarantee” property, which in turn, gets verified through analysis results. 

The CHESS methodology follows the “Correctness by Construction” practice which enforces \emph{(1)} the use of formal and precise tools and notations for the development and the verification of all product items; \emph{(2)} say things only once to avoid contradictions and repetitions; \emph{(3)} the design of software components that are easy to verify, by e.g., using safer language subsets, and to implement, by using appropriate coding styles and design
patterns.
 \cite{chessEmbedded}. 

\subsection{Multi-view modeling approach}

\begin{figure*}[!ht]
    \centering
    {\epsfig{file = 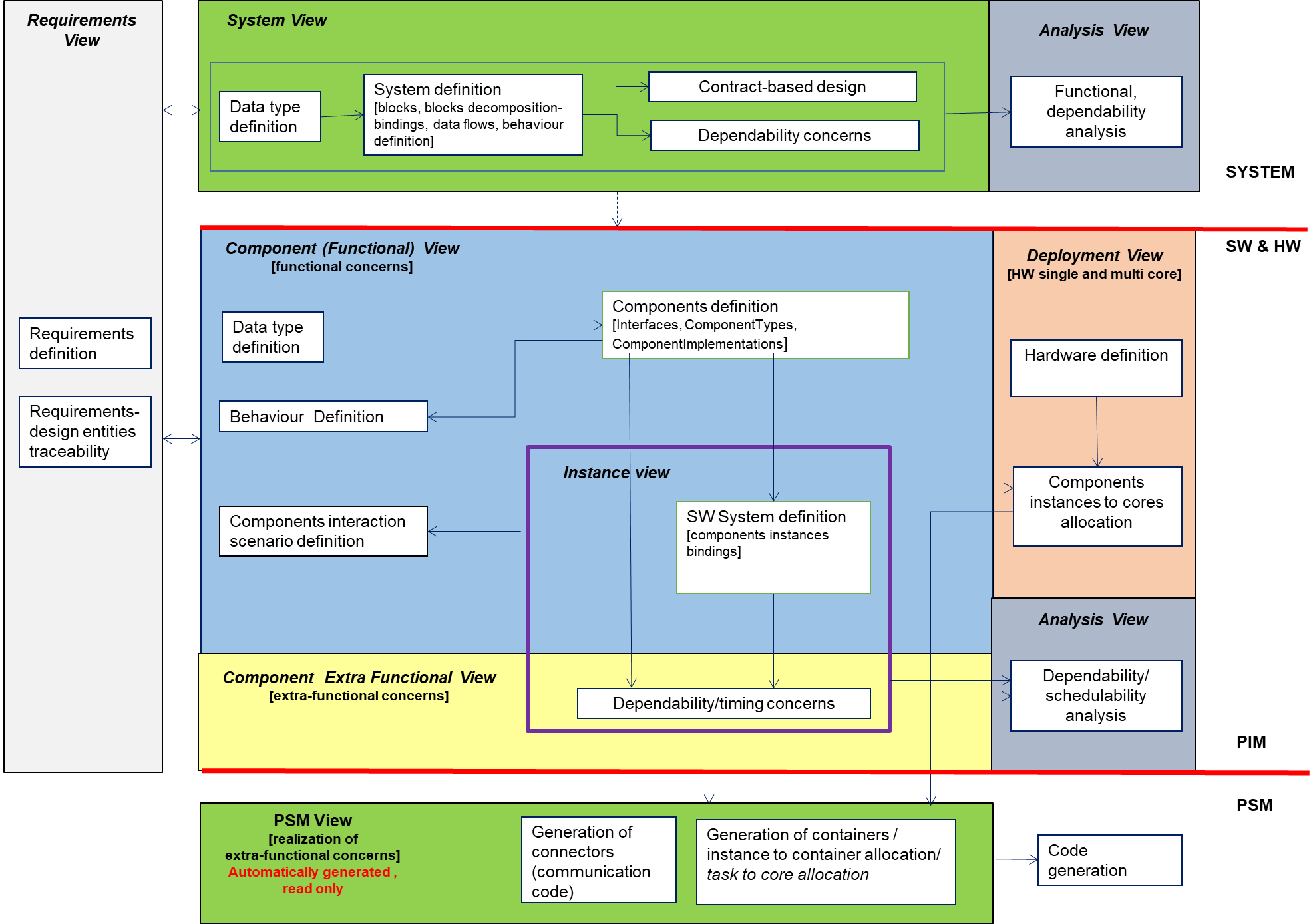,width=\textwidth}}
    \caption{CHESS views architecture,
    }
    \label{fig:chessviewArchitecture}
\end{figure*}

The CHESS tool provides a set of design views to uphold the "separation of concern", the "correctness by construction" and the other methodological principles introduced before. Six main views (requirement, component, system, deployment, analysis and instance views) are defined to support The CHESS modeling approach. Throughout the development process, each view has its own underlined constraints that enforce its specific privileges on model entities and properties that can be manipulated. Depending on the current stage of the design process, CHESS sub-views are adopted to enhance certain design properties or stages of the process. Figure \ref{fig:chessviewArchitecture} shows the high level architecture of CHESS views and their inter-relations.

\subsubsection{Requirement view}
Originally adopted from the SysML requirement diagram, the requirement view is used to define system requirements and track their verification. In CHESS, requirements are part of the model and play a central role in the system development life cycle. The system elements are associated with the technical requirements they satisfy, which are, in turn, traced to higher-level requirements, up to system-level requirements \cite{chessIoT}. This association technique enhances the traceability while evaluating the correctness and consistency of the modeled system. In this way, the change's impact can be better evaluated and faithful model verification evidence can be provided according to the requirements.

\subsubsection{System view}
It provides a suitable frame for system-level design activities. In the System view, the system entities are initially designed into blocks and then hierarchically decomposed. 
The system view supports contract-based design and several functional and dependability analysis. CHESSML inherits from SysML the specification of the block hierarchies and their internal decomposition, i.e. a block definition diagram can describe a system structure by means of a set of blocks and each block may have its own dedicated internal block diagram describing its sub-blocks decomposition and interfaces.

\subsubsection{Component view} This view is used for software design work and logic of the intended model. The component view is composed of two sub-views, \textbf{Functional View} which is enabled by default, and the \textbf{Extra-Functional View} which is enabled manually in the tool. The \textbf{Functional View} is used to model system functional specifications using diagrams such as class, composite structure, state machine, activity, and sequence diagrams. On the other hand, the \textbf{Extra-Functional View} is used to compose the system’s extra-functional specifications such as the real-time and dependability attributes. Recall that all views have a dedicated palette depending on their requirements, for instance, the extra-functional view has no access to the activity diagram and has the palette with  entries exclusively related to extra-functional concerns.
 
\subsubsection{Deployment view} This view is used to model for software design the hardware structure of the system and permits the allocation of their corresponding software component instances. Through the use of class and composite structure diagrams, the user can model the type of deployment on either single or multi-core processor. In this view, each hardware resource is allocated to a specific memory partition and can only access and change its own memory space. Regarding the software to hardware resources allocation, all software components are allocated to cores. 

\subsubsection{Analysis view} This is used to capture all the activities and diagrams related to analysis in CHESS. 
The analyses performed in CHESS are real-time analysis, quantitative dependability analysis, failure propagation analysis, and so on. We will discuss further on analysis in section \ref{analysis}.  
 
\subsubsection{Instance view}
CHESS provides a dedicated view to visualize and model the Platform Specific Model (PSM) as a combination of hardware and software instances generated from the deployment and component views respectively through the composite structure diagrams. This is a novel approach to facilitate the analysis between model instances. These instances are automatically generated when the BuildInstance command is invoked. In the generated instance model each component’s property and connector are mapped onto a dedicated \textit{InstanceSpecification}.

\subsection{Model-based Analysis and verification}\label{analysis}

CHESS provides the capability to perform several kinds of analysis
depending on the specific requirements (functional, timing,
dependability). Many of these functionalities have been added to the
new release of CHESS and are further explained in the next section.

\textit{\textbf{Functional Verification}} by means of model checking is supported by the integrated \emph{nuXmv} model checker~\cite{CavadaCDGMMMRT14}. System and component properties, derived from requirements, can be formalized into linear temporal logic properties, then they can be verified on top of the system's or component's behavioral models developed using state machines.

\textit{\textbf{Contract-Based Analysis}} is built on top of the OCRA tool support~\cite{CimattiDT13}. Component formal properties are structured in terms of contracts, comprised of an assumption and a guarantee pairs. The assumption is a restriction on the component’s environment or usage, and the guarantee is a property that must be satisfied by the component - provided that the environment satisfies the assumption \cite{chessMCritical}.

\textbf{\textit{Timing Analysis}} is built on top of the MAST\footnote{\url{https://mast.unican.es/}} analysis tool. It is invoked to perform analysis such as schedulability and end-to-end response time analysis. Schedulability analysis is performed by taking input from the annotated PSM model and the computed partition schedule on each available processing unit. Then, the response-time analysis calculates the worst-case response time of each task\cite{chessavionic} assessing the schedulable tasks complying with the given timing constraints. The end-to-end analysis is done by utilizing the component sequence diagram. Applying MARTE timing stereotypes, the tool evaluates the hardware component's responsiveness. This analysis facilitates “early end-to-end response time verification", giving a sense of any possible refinement of the model before deployment \cite{chessIoT}.

\textbf{\textit{Model validation}} exploits several types of methods to assess the software system on its target platform. We can mention:
(1) \textit{Model core constraints validation} is performed to enforce the CHESS model constraints including specific preconditions as required by the schedulability analysis.
(2) \textit{Validate model for state-based analysis},
(3) \textit{Validate model for model checking},
(4) \textit{Validate model for criticality specification} and finally
(5) \textit{Validate model for Automotive 26262 compliance (only specific for automotive domain)} checks the system correctness of Automotive Safety Integrity Level(ASIL) inheritance and decomposition according to the ISO 26262 standard.

\section{\uppercase{New System-Level Analysis Support}}\label{sec:updates}
In this section, we present the new major features released in CHESS 1.0.0. The new release includes the extension support for system-level safety and security analysis. CHESSML dependability profile which normally supports different techniques for safety and dependability analysis has been extended to model fault injection and threats. Other new features include contract validations, parameter-based architectures, and document generation. This new release can be accessed at \url{https://www.eclipse.org/chess/}

\subsection{Fault Injection and Model-Based Safety Analysis with xSAP}

The new release supports the conduct of Fault Tree Analysis (FTA) and Fault Mode Effect Analysis (FMEA). FTA is a deductive technique for identifying, evaluating, and modeling the interrelationship between events leading to a failure or an undesired state. FMEA is a highly structured approach through which all potential failure modes of a system and their effects can be identified, evaluated, and prioritized \cite{BV13}. 

Once the system model is defined in CHESS, through components definition and their nominal behavioral model, the faulty behavior is expressed through a specific state machine called "Error Model". The Error Model extends the nominal state machine with information about the effect upon a property of the component, and consequently on its nominal behavior. The next figure represents an example of an error model that, in case of an internal fault, moves the related component in an error state where the property "energy" is stuck at 0 value. The optional probability assigned to that transition is $5 x 10^{-2}$.
\begin{figure}[!h]
  \centering
   {\epsfig{file
   =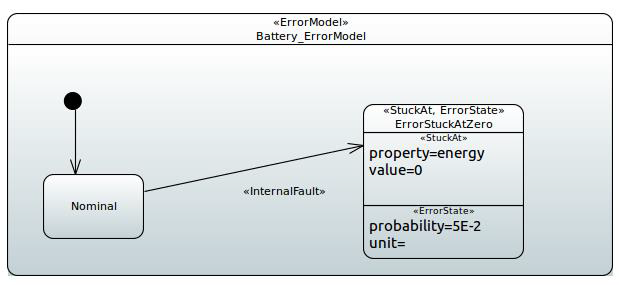,width=\columnwidth}}
    \caption{State machine modeling faulty behavior}
    \label{fig:errormodel}
\end{figure}
Once the error model is defined, the FTA or FMEA can be done by invoking the xSAP symbolic model checker through the CHESS environment. 

The xSAP approach is based on the library-based fault injection (i.e., an extension of a behavioral model with the definition of faults taken from a library of faults) and the use of model-based routines to generate safety artifacts.  

The result of the FTA is the fault tree that is automatically shown in a dedicated panel in the front-end; see Figure \ref{fig:errormodel} for an example. If fault probabilities have been specified during the configuration of the error model, the fault tree will report their combination. The fault tree can be examined for the minimal cut-set, identifying the basic fault conditions which can lead to the top-level failure.
\begin{figure}[!h]
  \centering
   {\epsfig{file =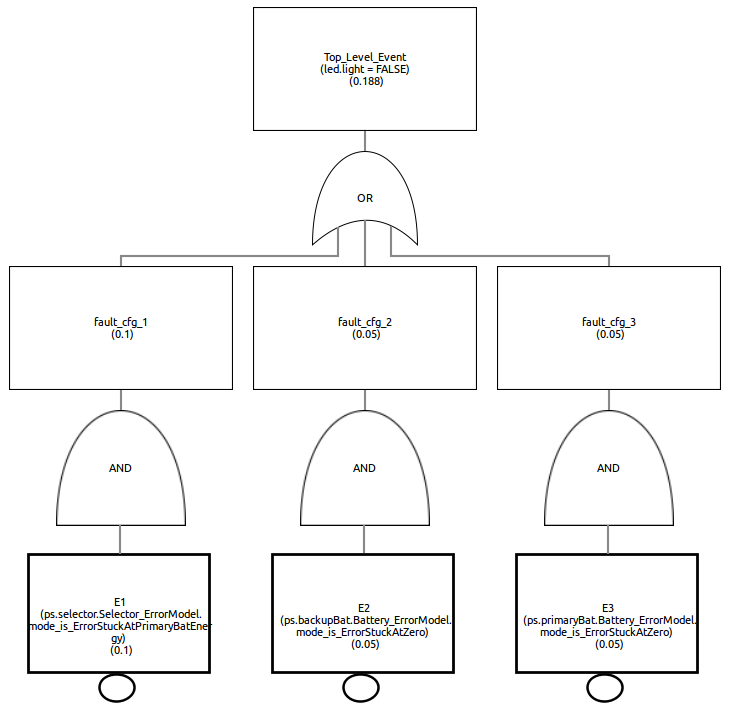,width=\columnwidth}}
    \caption{A fault tree visualized in the CHESS Editor View; note the probabilities associated to the top and basic events}
    \label{fig:faulttree}
\end{figure}
This is additional to the already existing analysis techniques in CHESS such as Failure Logic Analysis (CHESS-FLA) \cite{chessFLA} and State Based Quantitative Dependability Analysis (CHESS-SBA) \cite{chessSBA}.

\subsection{Reliability Analysis with Mobius}\label{mobius}
\textbf{Möbius}\footnote{\url{https://www.mobius.illinois.edu/}} is a software tool for modeling the behavior of complex systems, by allowing the study of the reliability, availability, security, and performance for large-scale discrete-event systems \cite{mobius}. Many reliability analysis results can be obtained with probabilistic models built with Mobius using the stochastic activity networks (SAN) formalism, solved via Monte-Carlo simulation\footnote{\url{https://www.investopedia.com/terms/m/montecarlosimulation.asp}}. The CHESS profile for dependability is used to enrich functional models of the system with information regarding the behavior with respect to faults and failures, thus allowing properties like reliability, availability, and safety to be documented and analyzed. 

The new release supports the modeling of security concerns which helps in threat identification at the early stages of the development and facilitates the exploiting of the Mobius capabilities for analysis of reliability. Specific extensions are related to the modelling of Cyber-Attacks aspects and models transformations from CHESS to the Mobius tool to run the analysis of SANs.

As a results the implemented a methodology allows modeling of a system security threat and a data corruption threat, which may result in a service corruption. An example of a system security threat can be a cyber-security attack, i.e. an unauthorized access of the system, halting services. 
\begin{figure}[!h]
  \centering
   {\epsfig{file =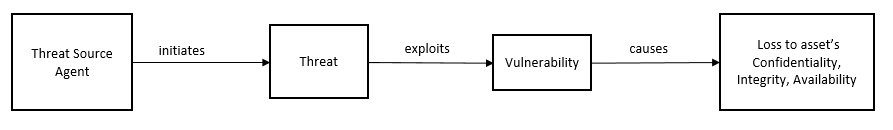,width=\columnwidth}}
    \caption{Process of Security breach }
    \label{fig:secBreach}
\end{figure}
Figure \ref{fig:secBreach} depict the process of a security breach that leads to the violation of the security-related properties. \textit{A threat} event, initiated by a threat source agent, able to exploit a \textit{vulnerability} of an asset (e.g. a component/system) may result in a loss to the confidentiality, integrity, and/or availability of the asset. Vulnerabilities could be represented as a pre-defined enumeration collected through different sources (e.g. personal competence, standards, results of previous threat analysis, etc.). Finally, the consequences could be modeled using pre-defined effects, which refers to the loss of Confidentiality, Integrity and Availability (CIA).
\begin{figure}[H]
    \centering
    \includegraphics[width=\columnwidth]{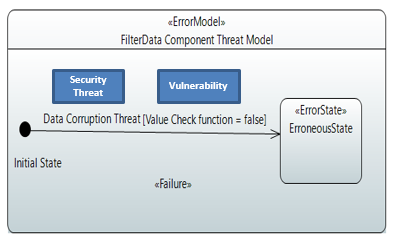}
    \caption{Erroneous state transition due to security threat event and vulnerability}
    \label{fig:errTransition}
\end{figure}
An {\textbf{\textit{\textless\textless ErrorModel\textgreater\textgreater}}}-tagged state machine is used when modeling the security breach. The failure, internal fault, and effect are extended to include security threats, vulnerability, and consequences respectively. Figure \ref{fig:errTransition} illustrates an example of an error model, where a cyber-security attack initiates a data corruption threat. The vulnerability was modeled exploiting the value check function which is set to false. In this case, the system transits to an erroneous state leading to component failure. Note that a component could have multiple instances of {\textbf{\textit{\textless\textless ErrorModel\textgreater\textgreater}}}-tagged state machines, attached to it. Each instance would provide the elaboration of input/output failure behavior addressing a specific concern.

The generation of the Mobius SAN model process is done by performing an automatic model-to-model transformation from a model instance to the SAN model recognized by Mobius for reliability analysis.

The new release facilitates the exploiting of the Mobius analysis capabilities for safety and security co-engineering needs, according to the scenario addressed envisaged in \cite{Popov_mobius}. Editing Mobius models cannot be trivial. To this purpose we have extended the CHESS profile and therefore the CHESS modelling language capabilities and user-friendly editor as front-end to fully support the modelling of system architectures taking in consideration safety and security co-engineering, and using automatic transformations to SAN model for reliability analysis with MOBIUS. This approach provides a smooth integration, guarantees the consistency among SysML and SAN models, and drastically reduce the effort required to construct an appropriate SAN analysis model. 

This extension has been developed in the context of the AQUAS project, as result of a collaborative effort among Intecs and City University of London, and applied across different use cases, in the  ATM and Industrial Drive domain.


\subsection{Improved support for contract-based design analysis and model checking} \label{imprSupport}

In the contract-based paradigm, the properties of each component may be restricted to its interfaces. The contracts are pairs of properties representing an assumption and a guarantee of the component. In addition, CHESS tool supports the contract refinement analysis for composite components. The contract of a composite component is defined by the assumption of the composite component itself and the guarantee ensured by the contracts of its sub-components, considering their interconnection as described by the architecture and and that the assumption of each sub-component is ensured by the contracts of the other sibling sub-components.

The new CHESS release has improved the contract-based analysis aspects by integrating CHESS with V\&V tools such as OCRA, nuXmv, and xSAP. In this regard, the new additional analysis includes:
 
(1) \textit{Model checking}, i.e. the behavioral models, that describe how the internal state of a component and the output ports are updated, can be verified against some formal properties in different temporal logics. The formal properties can represent some requirements (e.g., functional or safety-related requirements) or some validation queries such as the reachability of states. 

(2) \textit{Contract-based compositional verification of state machines} is performed on composite components. The local state machine of each sub-component is verified separately against its local contract. The correctness of the composite component is implicitly derived by the correctness of the contract refinement and the successful verification of local state machines. 

(3) \textit{Contract-based safety analysis}, i.e. identify the component failures as the failure of its implementation in satisfying the contract. When the component is composite, its failures can be caused by a failure of one or more sub-components and/or a failure of the environment in satisfying the assumption. As result, the analysis produces a fault tree in which each intermediate event represents the failure of a component or its environment, linked to a Boolean combination of other nodes. The top-level event is the failure of the system component. The basic events are the failures of the leaf sub-components, in addition to the failure of the environment (see \cite{bozzano2014formal} for more details).

\subsection{Support for parameterized architecture and trade-off analysis}
In a parameterized architecture the number of components, the number of ports, the connections, and the static attributes of components depends on a (possibly infinite) set of parameters.
In the new release, it is possible to define a parameterized architecture, setting the multiplicity of FlowPorts and sub-components to express a list of elements with the same type. This can be very helpful when modeling a system with a large number of similar nodes. 
The modeling of the parameterized architecture is followed by its instantiation. In this phase the user sets the values of the parameters, defining the configuration of the architecture.
OCRA takes in input the parameterized architecture and one or more configurations. Then, OCRA produces the instances of the architecture, and for each of them, performs a list of contract-based verifications. The output is the result derived from the contract-based verifications described in Section\ref{imprSupport}.

This new release also supports Tradeoff analysis which allows to execute a certain check such as safety, security, performance analysis on selected configurations (instances) and get the results in a view that simplifies the comparison between them. This makes it easy to visually get an idea of how the intended model instances perform with respect to the selected configurations.
\begin{figure}[!h]
  \centering
   {\epsfig{file =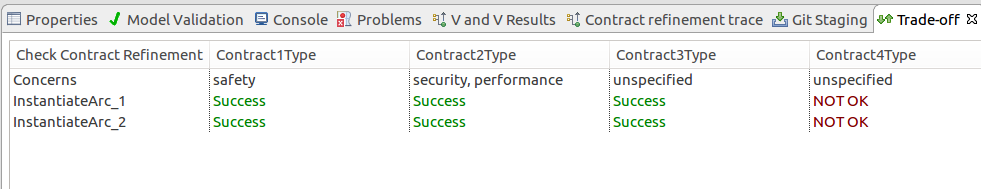,width=\columnwidth}}
    \caption{Trade-off Analysis results sample}
    \label{fig:tradeoff}
\end{figure}
Figure \ref{fig:tradeoff} shows the sample result of a trade-off analysis made on two instances by looking at different concerns specified by the assumption/guarantee formal properties of each contract.

\subsection{Automatic generation of diagrams and documentation}
The traditional way of editing a model is by adding an element in a diagram but changes made in the model are not reflected in the diagrams. The new CHESS release offers the possibility of generating a diagram from the model which reflects the data in the model on the fly. The supported diagrams are Block Definition Diagram(BDD) and Internal Block diagram(IBD). Multiple diagrams can be generated on a single component in the model. The generated diagram elements will be automatically aligned but the user can rearrange by moving elements manually or by invoking \textbf{“layout selection command.”}
 
\begin{figure}[!h]
  \centering
   {\epsfig{file =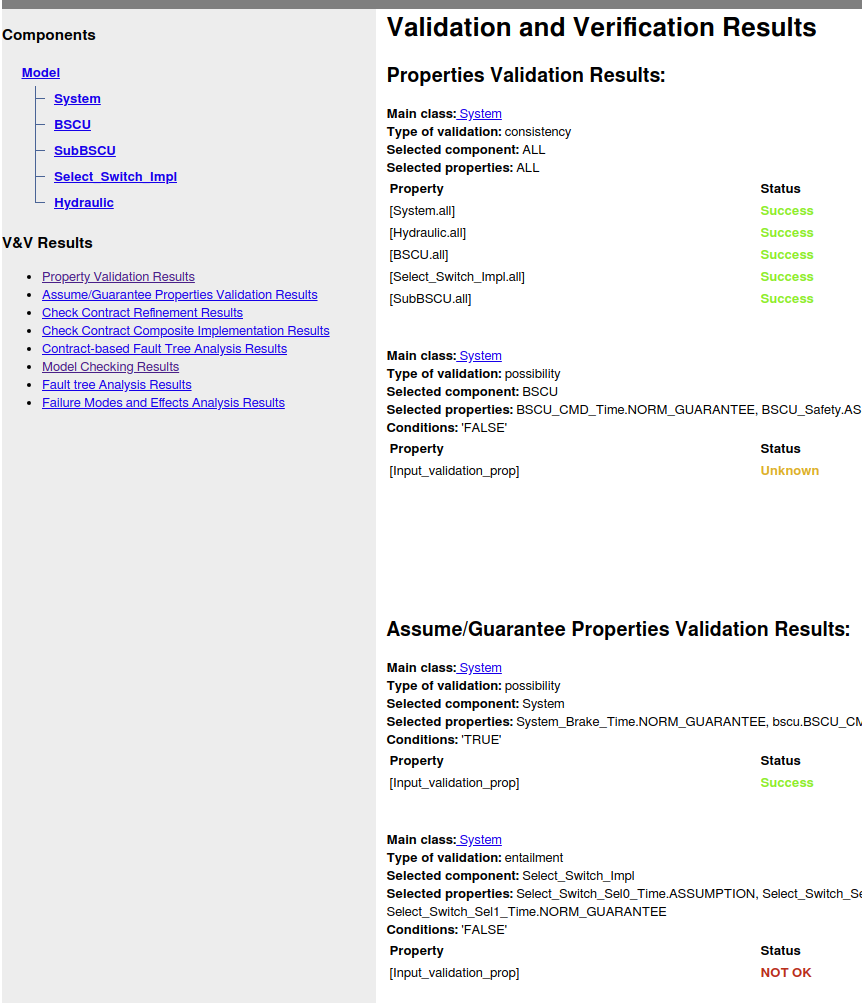,width=\columnwidth}}
    \caption{Generated report sample}
    \label{fig:report}
\end{figure}

The new release also supports the generation of the model architecture and the report on various analyses executed on the model in an HTML document or a LaTeX source code. The report is divided into two sections. The first describes the structure of the model which includes diagrams and the associated components while the second includes the report lists of the results of the validation and verification (V\&V) analyses results. There are many types of V\&V results such as \textit{property validation, assume/guarantee properties results, contract checks results, model checking, FTA, FMEA}, and so on. Figure \ref{fig:report} gives the sample of the report page that can be generated by the tool.

\section{\uppercase{CHESS Tool IN PRACTICE}}\label{sec:chessInPractice}

Throughout different phases of extending CHESS, CHESS was involved in many projects, research communities, and academies. CHESS has been used for teaching and extending it for research purposes. CHESS has conducted more than 10 international recognized research and development projects in the frame of 8 years\footnote{\url{https://projects.eclipse.org/projects/polarsys.chess/releases/1.0.0/review}}. Following we list a brief extent on the projects through which the CHESS tool played a major role.

\subsection{ARTEMIS JU-CHESS}\label{artemis}
ARTEMIS JU-CHESS\footnote{\url{http://www.chess-project.org/}} is the originating project that developed the CHESS tool. The project aimed to improve model-driven engineering practices and technologies to better address safety, reliability, performance, robustness, and other non-functional concerns. This was achieved while guaranteeing the correctness and composition of components under development in the embedded systems domain. From this point, the various project was initiated to extend CHESS to a new level by adding more functionalities and incorporate other domains.
 
 \subsection{CONCERTO}
CONCERTO\footnote{\url{http://www.concerto-project.org/}} project aimed to deliver a reference multi-domain architectural framework for complex, highly concurrent, and multi-core systems, where non-functional properties (including real-time, dependability, and energy management) was established for individual components.
CONCERTO framework was built on top of the CHESS framework developed in \ref{artemis}, as well as the results of several other related projects. The project enforced the modeling of multi-core processors among the possible target platforms, with the same level of correctness and guarantees as for traditional single-core processor targets \cite{lessonLerned}.

\subsection{AMASS}
AMASS\footnote{\url{https://www.amass-ecsel.eu/}} (Architecture-driven, Multi-concern and Seamless Assurance and Certification of Cyber-Physical Systems) project aim was to create and consolidate the de-facto European-wide open tool platform, ecosystem, and self-sustainable community for assurance and certification of Cyber-Physical Systems (CPS) in the largest industrial vertical markets \cite{amass1}. 
In this project, CHESS played a role in system architecture modeling assurance, patterns library management assurances, contract-based assurances, and verification and validation (V\&V) based assurances through its extension with some tools such as OCRA, nuXmv, and xSAP. For future consultation, the contributor offers free training accessible at \url{https://www.amass-ecsel.eu/content/training} which includes video tutorials and documentation.
\begin{figure}[t]
    \centering
    \includegraphics[width=\columnwidth]{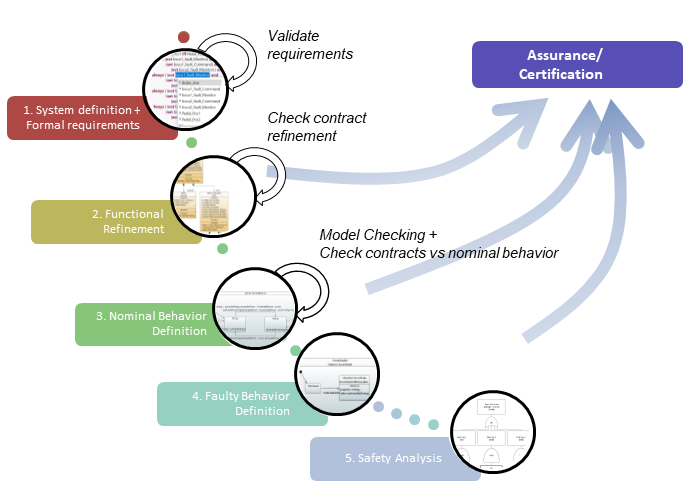}
    \caption{The main workflow supported by CHESS in AMASS}
    \label{fig:CHESSWorkflow}
\end{figure}
Figure \ref{fig:CHESSWorkflow} shows the main workflow supported by CHESS in the context of AMASS.  The workflow included the definition of the system and their formal requirements, the functional refinement of the system, the definition of the nominal and faulty behaviors, and finally, the safety analysis.

\subsection{AQUAS}
AQUAS\footnote{\url{https://aquas-project.eu/}}(Aggregated Quality Assurance in Systems) project aims was to improve on how the non-functional requirements of safety, security, performance (SSP) are dealt with during the product life cycle for embedded computer systems. AQUAS approach was based on two main principles. First, apply the methods for combined analyses of project artifacts from the viewpoints of safety, security, performance. Second, limit the overhead cost of these combined analyses by only applying them at a limited number of points in the product life-cycle, called interaction points \cite{D3.3.AQUAS}.

The main CHESS's contribution to the project was to support the modeling of the software architecture of the several use cases. CHESS was used to enrich the system and software architecture model with information related to safety, security and performance co-engineering, and support of automated toolchains and traceability for combined analysis among the product. In particular, the tool was used to perform performance analysis and support dependability  analyses with SANs, exploiting new capabilities presented in section \ref{mobius}.

\subsection{SESAMO}
SESAMO\footnote{\url{http://sesamo-project.eu/}} project aimed to develop a methodology to reduce interdependencies between safety and security mechanisms. This was achieved by constructing a tool-chain that uses the constructive elements and integrated analysis procedures to ensure the safety and security characteristics of the system are maintained. The CHESS contribution
to the project was (1) the introduction of the concept of components and reusability into the modelling process. (2) the definition of safety and security as non-functional properties, within the CHESS component model perspective.
(3) provide  separation of concerns for the modelling of safety and security in a single model enabling joint verification and analysis \cite{sesamoteam_2014}.

\subsection{MEGAMART}
MegaM@Rt\footnote{\url{https://megamart2-ecsel.eu/}} is an open-sourced project with the ambition to create a framework which incorporate methods and tools for continuous development and validation. This project leverages the advantages in scalable model-based methods to provide benefits in significantly improved productivity, quality, and predictability of large and complex industrial systems. 
\begin{figure}[!h]
  \centering
   {\epsfig{file =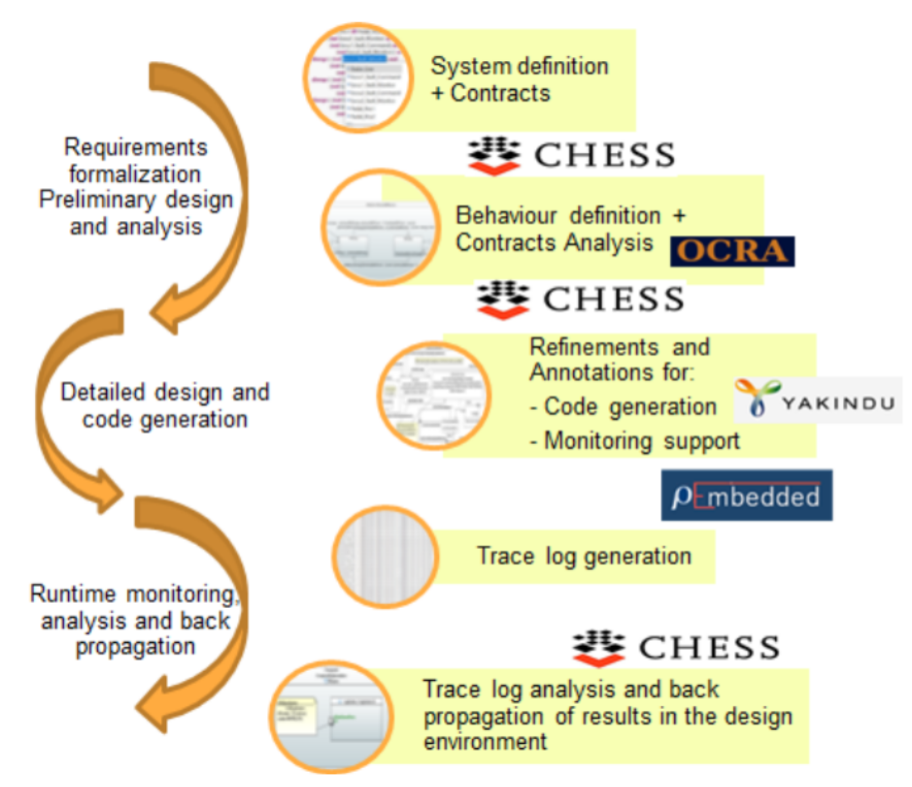,width=\columnwidth}}
    \caption{CHESS fit in MegaM@Rt Tekne case study}
    \label{fig:tekneCase}
\end{figure}
Employing its rich model-driven toolchain, CHESS served in the design and the development of high-integrity systems with a focus on non-functional properties\cite{chessMegamart}. In this project, CHESS was used in the design and analysis of the Tekne case study. This was an ultra-wideband (UWB) mobile network technology with a short-range communication, indoor positioning, and tracking capabilities \cite{chessTekne}.

Figure \ref{fig:tekneCase} shows the process of modeling the Tekne case study. In this project, requirement modeling, traceability, contract-based design approach, component real-time behavior analysis exploit and demonstrate the CHESS potential. In addition, the collected run-time logs were analyzed by CHESS to capture non-functional constraint violations. The results are back-propagated to the design environment for further model refinement \cite{chessTekne}.

\section{\uppercase{RELATED WORK}}\label{sec:relatedWork}

Several commercial tools provide similar functionalities of CHESS. One of the most popular is \textbf{Matlab/Simulink}\cite{simulink}. Although Matlab/Simulink facilitates the modelling and analysis of complex systems, its simulation efficiency might be an important disadvantage. Being based on a single Model of Computation and Communication (MoCC) is another limitation. \textbf{CoFluent} is another commercial tool extended to model IoT systems \cite{cofluent}. Although supporting more interaction models than Matlab/Simulink, it is also limited in the way components may interact among them. 

Another tendency is to overcome the UML lacks in semantic content, required in some application domains, towards a proliferation of DSLs \cite{bookmanuel}. Among the available DSLs, UML/MARTE is the standard language for real-time and embedded systems design, while SysML is the standard language for system modeling. Several modelling environments like \textbf{Papyrus} \cite{papyrus} support UML/MARTE. Nevertheless, its flexibility and semantic richness requires the definition of efficient modelling methodologies.

\textbf{Capella} \cite{capella} is an open-source comprehensive and extensible Eclipse system modelling tool. It is inspired to the SysML principles and it supports the \textbf{\textit{ARCADIA}} methodology that is successfully deployed in a wide variety of industrial contexts \cite{ARCADIA}. ARCADIA provides architectural descriptions for functional analysis, structural analysis, interfaces and behavior modeling, structured in five perspectives according to major system engineering activities and concerns.

\textbf{COMPASS}~\cite{BozzanoBCKNT19} supports model checking, model-based safety, reliability, and performance analysis and shares with CHESS some of the tools used as backend for such analyses. Differently from CHESS, it targets a variant of AADL and does not support traceability and code generation.

\textbf{MapleSim}\footnote{\url{https://www.maplesoft.com/products/maplesim/}} is a modeling tool for multi-domain engineering systems built on top of \textbf{\textit{Modelica}} modeling language \cite{modelica}. MapleSim features an integrated environment in which the system equations can be automatically generated and analyzed \cite{mapesim2}. 

Although we see some approaches able to tackle modeling challenges, no tool or approach has been able to fit in our methodology with such analysis and verification functionalities. Which makes CHESS a novel approach for implementing component-based modeling methodology for real-time and dependable systems by taking care of non-functional properties and enforces the correctness at all the stages of the development process.

\section{\uppercase{FUTURE WORK}} \label{sec:futureWork}
CHESS is a very huge toolset with more sophisticated and powerful functionality to meet user needs. However, there is still a gap for improvement, to cover more and more domains such as the Internet of Things (IoT) in a more concrete way.  Note that we are not concluding that it is not capable to perform some basic modeling of IoT related scenarios but we aspire to make it more IoT specific. This extension will follow CHESS’s component-based methodology and it will also follow already existing modeling approaches present in CHESS.
 
The envisioned approach will be achieved by improving the CHESSML metamodel with a set of specific stereotypes, contracts, communications, and operations profiled for IoT. The new proposed approach will also take in use of already existing dependability analysis infrastructure such as Fault Mode Effect Analysis, Fault Logic Analysis, Fault Tree Analysis, and so on. We also plan to export IoT models developed with CHESS to external consumers. Finally, we plan to exploit the current CHESS's code generation support for Ada language, integrated with the open-source ThingML framework, for IoT code generation. 






\section{\uppercase{Conclusions}}
\label{sec:conclusion}
Dependable complex system design and development present several challenges, the well-known canonical approach is to divide complex systems into smaller chunks (or subsystems), build them separately, and later integrate them. In this paper, we presented the current state of the CHESS tool to tackle design, analysis, and verification of real-time dependable complex systems. We walked through the CHESS tool architecture and we highlighted its component-based and multi-view modeling approaches. We have also presented the new system-level extensions and capabilities of the tool released under the CHESS1.0.0 version. Finally, we introduced the different projects and contributions where CHESS was used either in the industry and academia at large and presented the future envisioned extension strategy.

\section*{\uppercase{Acknowledgements}}
\noindent This work has received funding from the Lowcomote project under European Union’s Horizon 2020 research and innovation program under the Marie Skłodowska-Curie grant agreement n813884. We would like to acknowledge also different projects funding leading to the mature realization of CHESS which include the CHESS\footnote{\url{http://www.chess-project.org/}}, CONCERTO\footnote{\url{http://www.concerto-project.org/}}, SESAMO\footnote{\url{http://sesamo-project.eu/}} under ARTEMIS Joint Undertaking  initiative, and AMASS\footnote{\url{https://www.amass-ecsel.eu/}}, and AQUAS\footnote{\url{https://aquas-project.eu/}} under ECSEL Joint Undertaking initiative. We would like to acknowledge the main contributors to the development of the  CHESS toolset, in particular Stefano Puri, Nicholas Pacini, Luca Cristoforetti and Pietro Braghieri. Finally, we would like to acknowledge also Prof. Davide Di Ruscio for the assistance on drafting this paper.

\bibliographystyle{apalike}
{\small
\bibliography{Bibliography}}
\end{document}